# An origin in the local Universe for some short γ-ray bursts

N. R. Tanvir[1], R. Chapman[1], A. J. Levan[1] & R. S. Priddey[1]

[1]*Centre for Astrophysics Research, University of Hertfordshire, College Lane, Hatfield AL10 9AB, UK.*

**Gamma-ray bursts (GRBs) divide into two classes[1]: 'long', which typically have initial durations of $T_{90}>2$ s, and 'short', with durations of $T_{90}<2$ s (where $T_{90}$ is the time to detect 90% of the observed fluence). Long bursts, which on average have softer γ-ray spectra[2], are known to be associated with stellar core-collapse events—in some cases simultaneously producing powerful type Ic supernovae[3–5]. In contrast, the origin of short bursts has remained mysterious until recently. A subsecond intense 'spike' of γ-rays during a giant flare from the Galactic soft γ-ray repeater, SGR 1806–20, reopened an old debate over whether some short GRBs could be similar events seen in galaxies out to ~70 Mpc (refs 6–10; redshift $z\approx0.016$). Shortly after that, localizations of a few short GRBs (with optical afterglows detected in two cases[11,12]) have shown an apparent association with a variety of host galaxies at moderate redshifts[11–14]. Here we report a correlation between the locations of previously observed short bursts and the positions of galaxies in the local Universe, indicating that between 10 and 25 per cent of short GRBs originate at low redshifts ($z<0.025$).**

The satellite-based γ-ray detector CGRO/BATSE triggered on roughly 500 short-duration bursts during its nine-year lifetime. Unfortunately, the (1σ) positional uncertainties for these bursts were typically many degrees, giving limited information to help identify even their host galaxies, let alone their progenitors. A handful of short GRBs were localized to smaller error boxes by the Interplanetary Network and the Beppo-SAX and HETE-II satellites, but these only showed an absence of bright candidate host galaxies or afterglows[15,16]. Recently, however, afterglow detections for three short GRBs have associated them with galaxies at redshifts $z\approx0.2$: the X-ray afterglow of GRB 050509B was close to a bright elliptical galaxy at $z=0.225$, suggesting physical association[13,14]; GRB 050709 produced an optical afterglow locating it to a late-type galaxy at $z=0.16$ (ref. 11); and GRB 050724 exhibited an afterglow located in another elliptical galaxy at $z=0.257$ (ref. 12). The energetics of these bursts, and their association with a variety of host galaxies (including those with only old stellar populations) provide support for the view that some fraction of short GRBs arise from the coalescence of neutron-star/neutron-star (NS–NS) binaries[17–19].

On the other hand, the recent observation of the 'hypergiant' flare from SGR 1806–20 (refs 6, 7) re-ignited interest in the idea that some short bursts could be distant

SGR flares. Although previous giant flares were bright enough to have been seen by BATSE perhaps as far as the Virgo cluster, the SGR 1806–20 event would have been detectable out to several tens of megaparsecs, appearing very much like a short-hard GRB.

If even a proportion of short bursts originate in nearby galaxies, then despite poor localizations they may show a measurable spatial correlation with the positions of low-redshift galaxies. The PSCz galaxy redshift survey[20] makes an appropriate comparison data set for the all-sky BATSE data because, being IRAS-selected, it suffers less from incompleteness at low Galactic latitudes than other nearby redshift surveys (although other catalogues show similar results, as described in Supplementary Information).

We considered all 400 $T_{90}<2$ s bursts with localizations better than 10 degrees from the BATSE catalogue (4B(R); ref. 21 together with web supplement cited therein). A cross-correlation plot between these GRBs and the (1,070) PSCz galaxies with heliocentric recession velocities, $v$, $\leq 2,000$ km s$^{-1}$ is shown in Fig. 1. This sample includes most galaxies within about 25 Mpc, encompassing the local supercluster, and specifically the Virgo, Fornax and Ursa Major clusters. A clear positive correlation is revealed, which is even stronger when the galaxies are restricted to earlier morphological types (specifically Sbc and earlier; that is, T-type$\leq 4$). The figure also shows, as expected, that the long-duration bursts are uncorrelated with these galaxies.

However, the cross-correlation function is not ideal for this purpose as errors in different bins are not independent, and it makes no direct use of known BATSE instrumental characteristics. In order to optimize the search for a correlation signal, we compute $\Phi$, the sum of all burst-galaxy pairs weighted by the probability that they could be seen at the observed separation (or greater) if they were truly associated, and further weighted inversely by the burst error-circle size:

$$\Phi = \sum_{i}^{All\ bursts} \sum_{j}^{All\ galaxies} \frac{1}{\varepsilon_i} \int_{\theta_{ij}}^{\infty} \frac{1}{\sqrt{2\pi}\varepsilon_i} \exp\left(\frac{-\theta^2}{2\varepsilon_i^2}\right) d\theta$$

where $\theta_{ij}$ is the separation between the $i$th burst and the $j$th galaxy, and $\varepsilon_i$ is the error circle (statistical and systematic[22]) of the $i$th burst position. Further discussion of the BATSE instrumental characteristics[23] is provided in Supplementary Information.

To quantify the significance of the observed value of $\Phi$, we also compute $\Phi_0$, which is the mean of a large number of simulated random burst distributions (each with the same number of positions as the number of bursts under consideration, the same positional errors, and distributed on the sky according to the known BATSE sky exposure map[24]) correlated against the same set of (T-type$\leq 4$) PSCz galaxy positions. The spread of simulated results around $\Phi_0$ allows us to test the null hypothesis that there is no correlation between the positions of bursts and galaxies. This null hypothesis is rejected at the 99.9% level, confirming the indications of the cross-correlation function.

Next we attempt to estimate the proportion of bursts that are associated with nearby galaxies. To do this, we constructed many more artificial short-burst data sets, this time with both a random component and a component correlated with the galaxies from the PSCz galaxy catalogue with $v \leq 2,000$ km s$^{-1}$ and T-type$\leq 4$ (that is, with a 'host' selected at random from the catalogue and with positions smeared according to the real error circles from BATSE). Figure 2 shows that the observed signal could be explained by a correlated component representing between 6% and 12% of BATSE bursts (1$\sigma$ range). In fact, this is likely to be an underestimate, as our catalogue certainly does not contain all the galaxies in this volume, just the infrared-bright ones, and there remains a thin zone of avoidance around the Galactic plane, and a region unsurveyed by IRAS that the PSCz survey does not cover (amounting to about 16% of the sky).

Repeating this procedure in bins of recession velocity $v=2,000–5,000$ km s$^{-1}$ and $v=5,000–8,000$ km s$^{-1}$, we continue to find significant correlations (at nearly the 2$\sigma$ level in each bin). This provides strong confirmation that the correlation seen in the low-redshift bin was not simply a chance coincidence. The cumulative proportion of correlated bursts as a function of limiting cut-off velocity is shown in Fig. 3. Inevitably it becomes harder to detect correlations as the galaxy (and presumably detected burst) volume density decreases with increasing distance, and the angular size of large-scale structure projected on the sky also reduces. The range of percentage correlated bursts conservatively suggests that a total proportion of between 10% and 25% of BATSE bursts originate within ~100 Mpc.

Our results are broadly consistent (discussed further in Supplementary Information) with previously reported upper limits for the proportion of BATSE short bursts originating at low redshifts[6-10]. They also explain the intriguing finding that short-burst positions on the sky show evidence for a weak auto-correlation signal[25,26].

At first sight, the most likely explanation of our result is that we are detecting a low-redshift population of short bursts associated with SGRs. However, SGRs being the remnants of short-lived massive stars, it is then rather surprising that a stronger correlation is seen with earlier-type galaxies, which include some galaxies with little recent star formation. Furthermore, although a positive correlation signal is seen when restricting the burst sample to only those with $T_{90}<0.5$ s, a somewhat stronger signal is seen with the $T_{90}>0.5$ s short bursts, which is again surprising given that the spike from SGR1806−20 had a duration[6] of around 0.2 s.

An alternative possibility is that some or all of the low-redshift short bursts have the same progenitors as the recently discovered short-burst population at redshifts $z>0.1$. This would have the merit of simplicity, but as we show in Fig. 4, the higher-redshift bursts so far detected would have been rather brighter than any detected by BATSE if they had occurred at $z<0.02$. Thus a rather broad intrinsic luminosity function, perhaps comparable to that of the long-duration bursts, is required to accommodate reasonable numbers of both local and cosmological examples within the BATSE sample. A combination of differing progenitor masses and beaming could plausibly give rise to such a broad luminosity function for NS–NS binary mergers[27].

If both cosmological and local short-GRBs arise from NS–NS coalescence, then their association with at least intermediate-age and possibly old stellar populations is to be expected (owing to the inspiral lifetime of such binaries). Furthermore, the rate implied by our work of several bursts per year within 100 Mpc is in good agreement with recent calculations of the rate density of such events, based on the observed double neutron star population in the Milky Way[28]. The presence of such a large local population of NS–NS merger events would be encouraging for their detection prospects with upcoming gravity wave detectors.

**Supplementary Information** is linked to the online version of the paper at www.nature.com/nature.

**Acknowledgements** We acknowledge the use of the publicly available BATSE Current Burst database. We thank M. Briggs for advice on BATSE instrumental characteristics. N.R.T, A.J.L, and R.S.P acknowledge support from UK PPARC and RC acknowledges the support of a University of Hertfordshire studentship.

**Author Information** Reprints and permissions information is available at npg.nature.com/reprintsandpermissions. The authors declare no competing financial interests. Correspondence and requests for materials should be addressed to N.R.T. (nrt@star.herts.ac.uk).

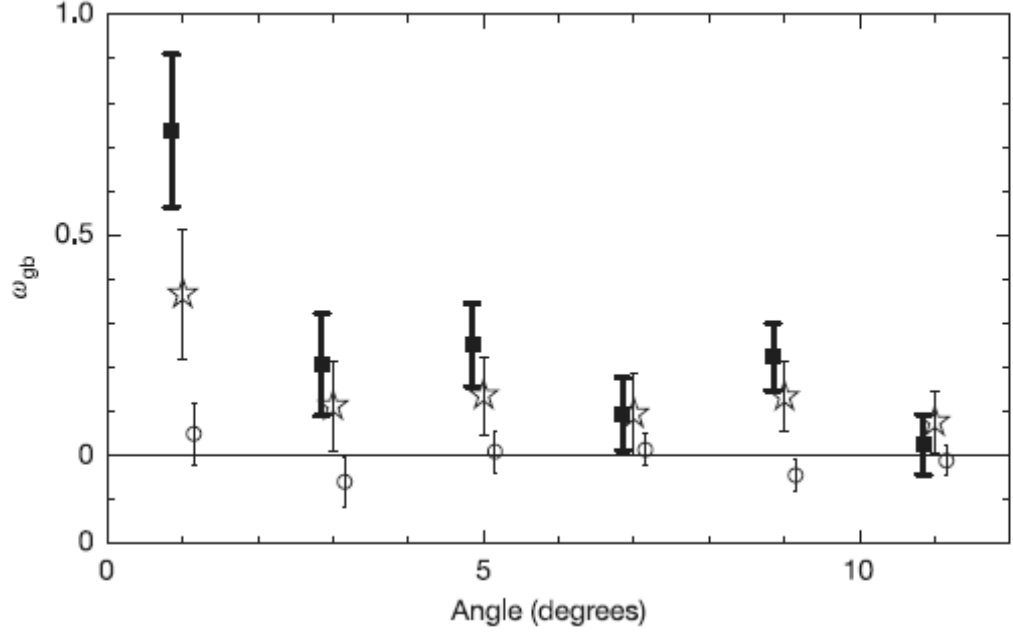

**Figure 1** Angular cross-correlation functionsfor BATSE short bursts with nearby galaxies from the PSCz catalogue. Two-point angular correlation functions, $\omega_{gb}$, (in two-degree bins) are shown for 400 BATSE short bursts ($T_{90}<2$ s and statistical position uncertainty ≤10 degrees) with nearby galaxies from the PSCz catalogue (recession velocities $v≤2,000$ km s$^{-1}$≈28 Mpc distant). Open stars show the correlation with all galaxy types (1,072 galaxies), and bold square symbols represent the same function but with galaxies restricted to earlier morphological types (T-type ≤4, Sbc and earlier, 709 galaxies). Open circles show the same function for long-duration bursts compared with all galaxy types. Points from the different functions have been offset slightly in angle for clarity. The short bursts exhibit clear (>3$\sigma$) correlation at low angles with the earlier-type galaxies, and a >2$\sigma$ correlation with all galaxy types. In contrast, as would be expected, the long bursts (1,481 events) show no measurable correlation with local galaxies. The highly anisotropic distribution on the sky of galaxies within this region makes it possible to detect this correlation signal. Uncertainties (error bars represent 1$\sigma$) in each bin were determined from Monte Carlo simulations, but true errors are not independent from bin to bin.

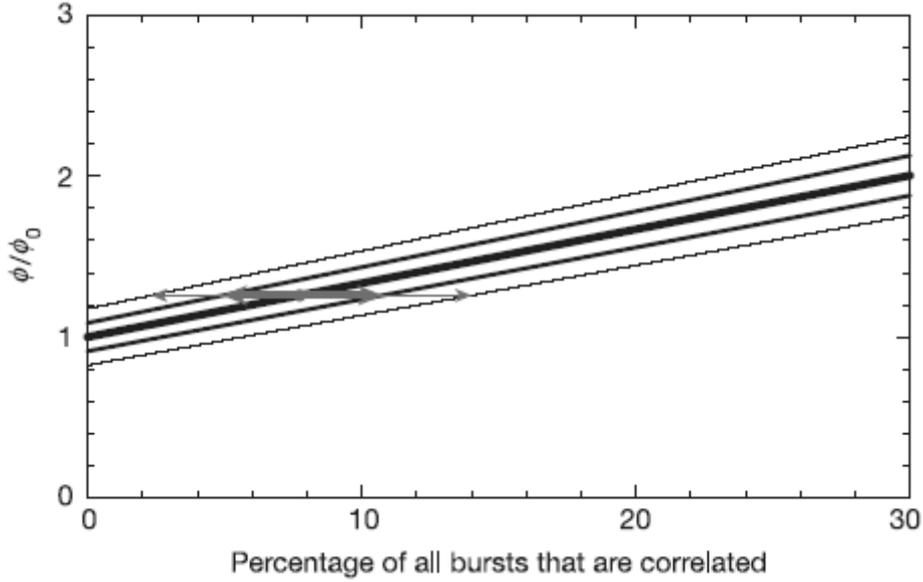

**Figure 2** Proportion of short bursts correlated with galaxies within $v \leq 2,000$ km s$^{-1}$. This figure summarizes an analysis of many simulated BATSE short-burst samples in which a proportion of bursts are laid down correlated with the PSCz galaxy positions and the remainder are placed at random. In this case, the galaxy sample is restricted to morphological T-type$\leq 4$ and within recession velocity $v=2,000$ km s$^{-1}$. The bold diagonal line represents the value of $\Phi/\Phi_0$ (see main text) as a function of the proportion of bursts in the simulated sample whose positions were seeded by the galaxy positions (random errors based on those in the BATSE catalogue were used to find offsets from these seed positions). The thinner lines show the $1\sigma$ and $2\sigma$ deviations around this line according to the simulations. The level of $\Phi/\Phi_0$ for the real data is illustrated by the horizontal arrow, which spans the $2\sigma$ (small arrow) and $1\sigma$ (large arrow) ranges. Thus the possibility of no correlation is rejected at more than the $3\sigma$ level, and a correlated fraction around ~8% is indicated (95% confidence limits ~2% to ~14%). As discussed in the text, there are reasons for expecting this to be an underestimate. Note that using all T-types produces a consistent, but less well constrained result.

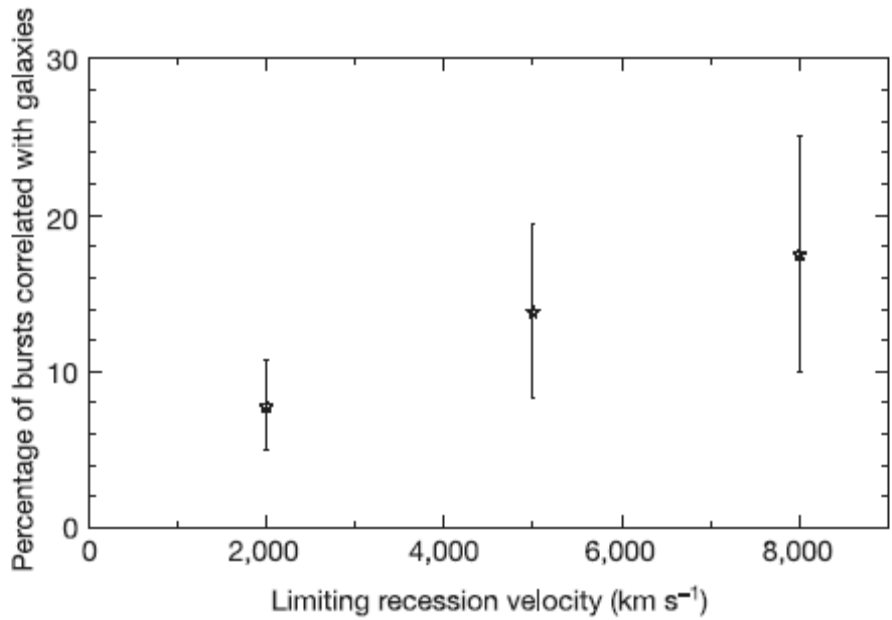

**Figure 3** The percentage of correlated bursts with increasing redshift. This figure shows the results of repeating the analysis of Fig. 2 for two further velocity-limited samples. Both the expected number of correlated bursts and $1\sigma$ ranges are shown. We emphasize that these figures are arrived at by comparing the relative values of $\Phi/\Phi_0$ for many simulated burst data sets with the real galaxy data set in each case. At higher redshifts the implied proportion of correlated bursts increases, as we would expect, but the statistical constraints also become weaker.

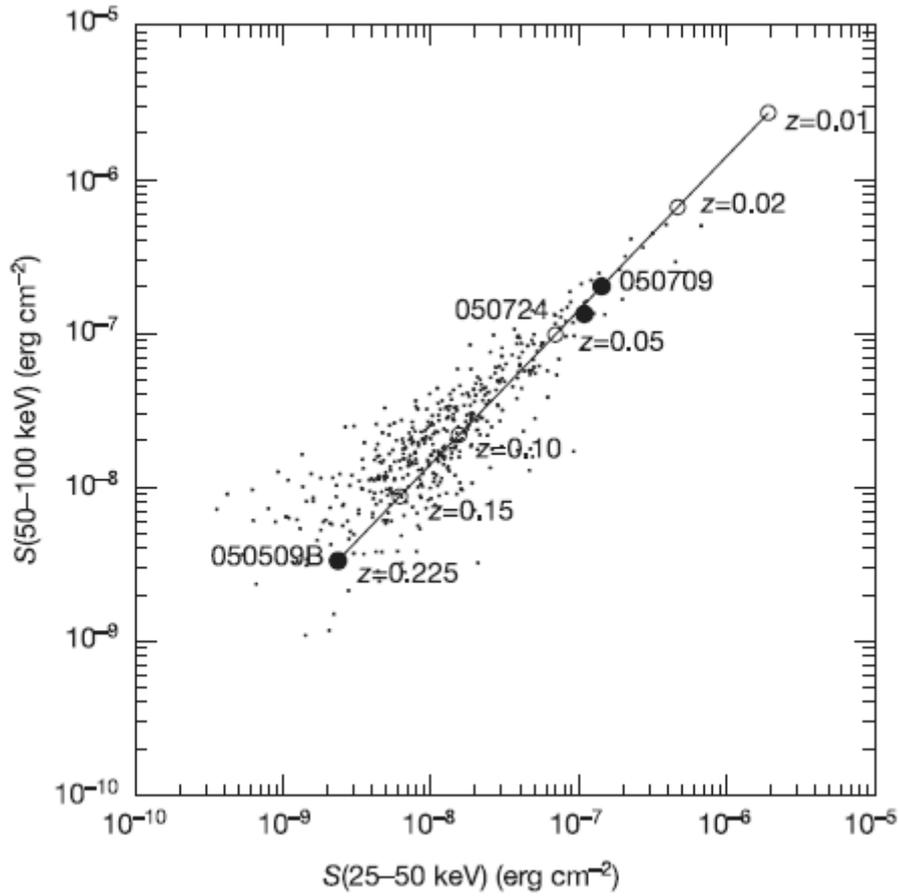

**Figure 4** The positions of GRB 050509B, GRB050709 and GRB050724 (bold points) in the BATSE fluence distribution. The measured γ-ray fluence ($S$) of GRB 050509b (ref. 14) was very low ($9.5\times10^{-9}$ erg cm$^{-2}$ in the 15–150 keV band) and, when converted into the full BATSE passband (25–300 keV), is essentially the faintest burst within this catalogue. As illustrated by the open points, if this burst had occurred within 100 Mpc it would lie in the brightest 1% of bursts, although it would be brighter than any BATSE bursts were it closer than 30 Mpc—the volume in which we measure our most significant correlation. The other two recently claimed short-burst identifications are brighter[29,30], and therefore if they had occurred at lower redshift would certainly have been the brightest in the BATSE sample. This implies that the bursts responsible for the correlation we measure are most probably intrinsically much less luminous than these three, although, as discussed in the text, a broad luminosity function might accommodate both with the same class of progenitor.

**SUPPLEMENTARY FIGURES**

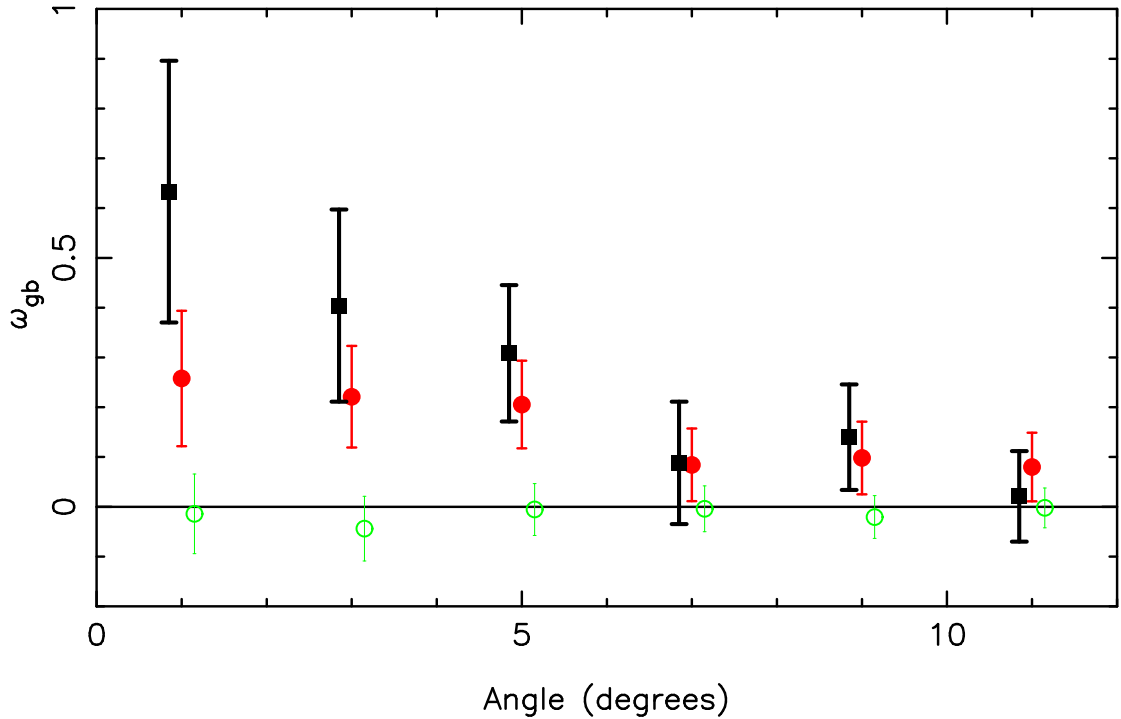

Supplementary Figure 1. Angular cross-correlation functions for BATSE short bursts with nearby galaxies from the RC3 catalogue

Two point angular correlation functions (in two degree bins) are shown for 400 BATSE short bursts ($T_{90}<2s$ and statistical position uncertainty ≤ 10 degrees) with nearby galaxies from the RC3 catalogue (recession velocities v ≤ 2000 km/s corresponding to roughly 28 Mpc In distance). Filled circles show the correlation with all galaxy types (~2300 galaxies), and bold square symbols represent the same function but with galaxies restricted to earlier morphological types (T-type <=4, Sbc and earlier, ~800 galaxies). Open circles show the same function for long duration bursts compared with all galaxy types. Points from the different functions have been offset slightly in angle for clarity. Again the short bursts exhibit clear correlation at low angles with the earlier-type galaxies, and correlation with all galaxy types. In contrast, as expected, the long bursts (1481 events) show no measurable correlation with local galaxies.

# SUPPLEMENTARY NOTES

## Comparison with another galaxy redshift catalogue

Although improbable, one can ask whether the observed correlations could be a chance consequence of the particular galaxy catalogue we chose. To test this we repeated the analysis against the large compilation of redshifts contained in the Third Reference Catalogue of Bright Galaxies[1]. Although there is some overlap, this compilation contains more galaxies, particularly early-types (but, of course, is more heterogeneous and biased against low galactic latitudes than is the PSCz). Supplementary figure 1 shows the correlation functions in this case, confirming that a statistically significant correlation is also seen here, albeit at slightly lower level of significance.

## Further discussion of the robustness of the statistic used.

Although our particular choice of is likelihood function (made a priori) is a reasonable approximation to an optimal scheme, it is true that the BATSE error circles are not strictly circular Gaussians, and that the systematic uncertainties are hard to characterise precisely (in practice we used error circles obtained by summing in quadrature the quoted statistical errors with a systematic error described by model 1 of Briggs et al.[2]). Moreover, the intrinsic distribution of short bursts around their hosts, which ideally should also be accounted for, is presently unknown. However, reassuringly, the results we discuss do not depend strongly on the precise scheme adopted. For example, arbitrarily increasing all the statistical errors in the BATSE catalogue by 50% only reduces the rejection significance with respect to the T-type<=4 PSCz (v<2000km/s) galaxies from about $3.3\sigma$ to $3.1\sigma$, whilst decreasing them all by 33% increases the rejection significance to $3.6\sigma$. Furthermore using other centrally peaked weighting functions apart from Gaussian also produces similar results.

1. De Vaucouleurs, G., de Vaucouleurs, A., Corwin, H.G., Buta, R.J., Paturel, G., Fouque, P., Third Reference Cat. of Bright Galaxies (RC3), Springer-Verlag: New York, (1991)

2. Briggs, M. S. *et al.* The error distribution of BATSE gamma-ray burst locations. *Astrophys. J.* **122,** 503–518 (1999).

**SUPPLEMENTARY DISCUSSION**

**Further comparison with previous work**

This work presents the first published correlation analysis specifically for short duration bursts and the all-sky distribution of low-redshift galaxies. As explained in the main text, there have been claims of an autocorrelation signal in the positions of BATSE short-GRBs[1,2], which is now readily explained by a proportion of bursts tracing nearby galaxies which are themselves clustered. Popov and Stern[3] specifically found no significant excess of short duration (T<0.7s) GRBs from the direction of the Virgo cluster. We confirm this finding, which is somewhat surprising, although if only about 30 BATSE bursts are from galaxies within v=2000 km/s, as indicated by our results, then the excess from Virgo galaxies in the PSCz should only be about 3 bursts (T90<2s), which is within the noise of the small number statistics.

1. Balazs, L.G., Meszaros, A., Horvath, I, Anisotropy of the sky distribution of gamma-ray bursts, A&A, **339**, 1-6 (1998)

2. Magliocchetti, M., Ghirlanda, G., Celotti, A, Evidence for anisotropy in the distribution of short-lived gamma-ray bursts, MNRAS, **343**, 255-258 (2003)

3. Popov, S.B., Stern, B.E., Soft gamma repeaters outside the Local group, submitted to MNRAS, astro-ph/0503532